\documentclass[aps,prd,superscriptaddress]{revtex4}

\usepackage{amsfonts}

\usepackage{amsmath}
\usepackage{graphicx}
\usepackage{color}
\usepackage{mathtools,leftidx}

\usepackage{soul}

\DeclareMathAlphabet\mathbfcal{OMS}{cmsy}{b}{n}

\begin{document}

\title{Casimir effect in Lorentz-violating scalar field theory: a local approach}

\author{C. A. Escobar}
\email{carlos_escobar@fisica.unam.mx}
\affiliation{Instituto de F\'{i}sica, Universidad Nacional Aut\'{o}noma de M\'{e}xico, Apartado Postal 20-364, Ciudad de M\'{e}xico 01000, M\'{e}xico}

\author{Leonardo Medel}
\email{leonardo.medel@correo.nucleares.unam.mx}
\affiliation{Instituto de Ciencias Nucleares, Universidad Nacional Aut\'{o}noma de M\'{e}xico, 04510 Ciudad de M\'{e}xico, M\'{e}xico}

\author{A. Mart\'{i}n-Ruiz}
\email{alberto.martin@nucleares.unam.mx}
\affiliation{Instituto de Ciencias Nucleares, Universidad Nacional Aut\'{o}noma de M\'{e}xico, 04510 Ciudad de M\'{e}xico, M\'{e}xico}

\begin{abstract}
We study the Casimir effect in the classical geometry of two parallel conductive plates, separated by a distance $L$, for a Lorentz-breaking extension of the scalar field theory. The Lorentz-violating part of the theory is characterized by the term $\lambda \left( u \cdot \partial \phi \right )^{2}$, where the parameter $\lambda$ and the background four-vector $u ^{\mu}$ codify Lorentz symmetry violation. We use Green's function techniques to study the local behavior of the vacuum stress-energy tensor in the region between the plates. Closed analytical expressions are obtained for the Casimir energy and pressure. We show that the energy density $\mathcal{E} _{C}$ (and hence the pressure) can be expressed in terms of the Lorentz-invariant energy density $\mathcal{E} _{0}$ as follows
\begin{align}
    \mathcal{E} _{C} (L) = \sqrt{\frac{1-\lambda u _{n} ^{2}}{1 + \lambda u ^{2}}} \mathcal{E} _{0} (\tilde{L}) , \notag 
\end{align}
where $\tilde{L} = L / \sqrt{1-\lambda u _{n} ^{2}}$ is a rescaled plate-to-plate separation and $u _{n}$ is the component of $\vec{{u}}$ along the normal to the plates. As usual, divergences of the local Casimir energy do not contribute to the pressure.
\end{abstract}

\maketitle

\section{Introduction}

Lorentz symmetry breaking has attracted great attention in the last decades both from the theoretical and experimental sides. This interest is justified from the fact that diverse quantum gravity candidates, such as loop quantum gravity and string theory, predict the breakdown of such fundamental symmetry at very short distances, presumably at the Planck scale. If Lorentz symmetry is really broken at very high energies, the effects of this violation should manifest at lower energies; however no violation has been detected so far. This is why the most important direction in the study of Lorentz symmetry breaking has been through low-energy effective field theories. Some well known Lorentz-breaking field theories are, for example, noncommutative field theories \cite{Mocioiu, Carroll, Carlson, Anisimov}, brane world scenarios \cite{Burgess, Frey, Cline} and the Standard-Model Extension (SME) \cite{Kostelecky, Kostelecky2}. Indeed, the latter has grabbed the most attention in the context of Lorentz violation. The SME allows a spontaneous violation of Lorentz symmetry, implemented through the emergence of nonzero vacuum expectation values of some vector and tensor fields, generating thus preferential directions in spacetime. Of course, this anisotropy in the spacetime should manifest as small deviations in any physically measurable quantity predicted by the Lorentz-symmetric theory.

The Casimir effect, predicted by H. B. Casimir in 1948 \cite{Casimir} and experimentally confirmed by M. J. Sparnnaay in 1958 \cite{Sparnnaay}, is one of the most remarkable consequences of the nonzero vacuum energy predicted by quantum field theory. In general, it refers to the stress on bounding surfaces when a quantum field (whether fermionic or bosonic) is confined to a finite volume of space. This force is due entirely to the change, brought about the presence of boundaries, in the energy of the vacuum. The relevance of the Casimir effect is apparent in many branches of physics, ranging from quantum field theory and theories with compactified extra dimensions \cite{Poppenhaeger, Edery, Cheng}, to gravitation \cite{Quach, Santos&Khanna-Grav, Jiawei} and condensed matter \cite{Cortijo, Cortijo2, MUC}.

It is worth mentioning that any effective field theory has an analogue of the Casimir effect, since it supports field oscillations as well. For example, the electromagnetic response of the topological phases of matter is described by effective electromagnetic field theories (which are obtained by integrating-out fermions in the microscopic Hamiltonian). So when we put two of these materials close each other, the zero-point energy of the field will be modified (as compared with the trivial electromagnetic vacuum), and hence a Casimir effect takes place. Regarding Lorentz-violating effective field theories, the Casimir effect has also been extensively studied, since the broken symmetry (which manifests through background vector and tensor fields) affects the Casimir force on bounding surfaces. For example, in the context of the Standard-Model Extension, the CE has been discussed within the electromagnetic \cite{Kharlanov, CE&AM, CE&AM2}, fermionic \cite{Frank&Turan, Santos&Khanna, Cruz&PetrovFermion} and gravitational \cite{Santos&Khanna2, Blasone} sectors.

In this paper we study the Casimir effect between two parallel conductive plates for a Lorentz-violating massive scalar field. The theory is defined by the Klein-Gordon Lagrangian plus the Lorentz-violating term $\lambda \left(u \cdot \partial \phi \right) ^{2}$, where the parameter $\lambda$ and the background four-vector $u ^{\mu} =(u ^{0}, \vec{u})$ control Lorentz symmetry breaking \cite{Gomes&Petrov}. Here, we employ a field theoretical approach to evaluate the vacuum expectation value of the stress-energy tensor (by means of the usual point-splitting technique and expressing it in terms of the corresponding Green's function), from which we calculate the Casimir energy (and pressure) in an analytical fashion. We find that the energy and pressure in the presence of Lorentz violation can be expressed in simple forms in terms of the Lorentz symmetric results. Remarkably, the results reported in Ref. \cite{Cruz&Petrov}, where the CE was studied by summing over the zero-point modes, are just particular cases of the results we present here. Furthermore, we also provide additional information about the divergence of the local energy density near the plates and provide a full expression for the vacuum stress.

The outline of this work is the following. In Sec. \ref{Intro} we present the theoretical model we deal with: a Lorentz-violating massive real scalar field. Section \ref{GFsection} is devoted to the derivation of the different Green's functions to be used within the local approach to the CE. Using the standard point-splitting technique of quantum field theory, in Sec. \ref{VSsection} we introduce the vacuum stress-energy tensor. The Casimir effect is fully analyzed in Sec. \ref{CEsection}. In Sec. \ref{Conclusection} we summarize our results and give further concluding remarks. Throughout the paper, natural units are assumed ($\hbar = c = 1$) and the metric signature will be taken as $(+,-,-,-)$.

\section{Lorentz-violating scalar field theory} \label{Intro}

Our starting point is the Lorentz-violating (LV) Lagrangian for a massive scalar field theory given by \cite{Gomes&Petrov}
\begin{align}
\mathcal{L} = \frac{1}{2} \left[ \left( \partial \phi \right)^{2} + \lambda \left(u \cdot \partial \phi \right)^{2} - m ^{2} \phi ^{2} \right],
\label{Lagrangian1}
\end{align}
where the second term encodes the breakdown of Lorentz symmetry. The constant vector $u ^{\mu} = (u ^{0}, \vec{u})$, which acts as a background field, does not transform under active Lorentz transformations. This vector, together with the dimensionless parameter $\lambda$, characterize the Lorentz symmetry violation. An important note, Lorentz-violating coefficients are usually assumed to be small and then a perturbative treatment is appropriate to see the effects of Lorentz violation in a given physical phenomena. In this paper we relax this assumption by considering that $k _{\mu \nu} \equiv \lambda u _{\mu} u _{\nu}$ has a finite value, not necessarily much smaller than 1. However, formal restrictions on $ \vert k _{\mu \nu} \vert$ can be derived from the positive-energy condition \cite{Gomes&Petrov}. Indeed, from the Lagrangian (\ref{Lagrangian1}), one can see that the regime $\vert k _{\mu \nu} \vert > 1$ produces instabilities in the field theory, since the kinetic term for motion flips its sign. This means that excitations with large momenta have lower energies and hence there is no vacuum state. In this paper we deal with the Casimir effect, which is a manifestation of the quantum vacuum. As such, we restrict ourselves to the field theory defined by the Lagrangian density (\ref{Lagrangian1}) in the regime $\vert k _{\mu \nu} \vert < 1$. 

It is worth mentioning that, in the limit $\vert k _{\mu \nu} \vert \ll 1$ and working to first order in $k _{\mu \nu}$, the Lorentz-violating scalar field theory can be actually transformed into the Lorentz-invariant theory by performing a suitable change of spacetime coordinates, i.e. $x ^{\prime \, \mu} = x ^{\mu} - \frac{1}{2} k ^{\mu} _{\phantom{\mu} \nu} x ^{\nu}$ \cite{Altschul1,Altschul2}. In this scenario, the Lorentz-breaking term in the Lagrangian (\ref{Lagrangian1}) can be eliminated, i.e. $\int d ^{4}x \, \mathcal{L} (\phi , \partial \phi) = \int d ^{4}x ^{\prime} \mathcal{L} ^{\prime} (\phi ^{\prime} , \partial ^{\prime} \phi ^{\prime})$, where $\mathcal{L} ^{\prime} = \frac{1}{2} [ \left( \partial ^{\prime} \phi ^{\prime} \right)^{2} - m ^{2} \phi ^{\prime \, 2} ]$ and $\phi ^{\prime} =  J ^{1/4} \phi$, being $J =  \vert \det (\frac{\partial x^\mu}{\partial x^{\prime\nu} }) \vert$ the Jacobian of the transformation. The coordinate redefinition method has proven to be useful in many Lorentz-violating field theories. It can be used to move Lorentz violation from one sector to another in interacting field theories, for example, from the fermion to the photon sector of the Standard-Model Extension. However, when the LV coefficient $k _{\mu \nu}$ is not much smaller than one, the coordinate redefinition method does not provide a good approximation and hence we have to work directly with the full Lagrangian (\ref{Lagrangian1}). This is precisely the case we consider in this paper.

The equation of motion arising from the Lagrangian (\ref{Lagrangian1}) reads
\begin{align}
\left[ \Box + \lambda \left( u \cdot \partial \right) ^{2} + m ^{2} \right] \phi (x)  = 0 , \label{EQLV}
\end{align}
and the stress-energy tensor for this theory is given by
\begin{align}
T ^{\mu \nu} = (\partial ^{\mu} \phi ) (\partial ^{\nu} \phi ) + \lambda u ^{\mu} (\partial ^{\nu} \phi ) (u \cdot \partial \phi) -  \eta ^{\mu \nu} \mathcal{L} . \label{Tmunu}
\end{align}
Here $\eta ^{\mu\nu} = \textrm{diag} (1,-1,-1,-1)$ is the usual Minkowski flat spacetime metric. Note that, unlike most of the standard cases where Lorentz symmetry is preserved, this tensor can not be symmetrized because its antisymmetric part 
\begin{align}
T ^{(\mu \nu)} = \frac{\lambda}{2}[u^\mu(\partial^\nu \phi)-u^\nu(\partial^\mu \phi)]\left(u \cdot \partial \phi \right)
\end{align}
is no longer a total derivative. We can directly verify that the stress-energy tensor (\ref{Tmunu}) is conserved, i.e. $\partial_\mu T^{\mu\nu}=0$; however, is not traceless $T^{\mu}\,_\mu\neq0$.

In this paper we are concerned with the Casimir effect associated with a Lorentz-violating scalar field theory confined between two parallel plates. To this end, we employ a local approach consisting in the evaluation of the vacuum expectation value of the stress-energy tensor (\ref{Tmunu}), which can be expressed in terms of the appropriate Green's function for the modified field equation (\ref{EQLV}).

\section{Green's function}
\label{GFsection}

Let us consider a Lorentz-violating massive scalar field confined between two large parallel plates separated by a distance $L$. In the following we derive the Green's function (GF) for the confined LV scalar field by imposing Dirichlet conditions on the plates. We will also derive another Green's functions relevant for the calculation of the renormalized Casimir energy and stress. The GFs for the case of Neumann or Robin boundary conditions can be derived in the same way.

For simplicity, we orient the coordinate frame so that one plate is at $z=0$ while the other is at $z = L$. So the unit normal to the former plate is $n ^{\mu} = (0,0,0,1)$. Also, we conveniently decompose the spatial part of the background LV 4-vector $u ^{\mu}$, $\vec{u}$, into the transverse $\vec{u} _{\perp}$ (along the $x$ and $y$ directions) and longitudinal $u _{z}$ (along the $z$ direction) components. In this way $u ^{\mu} = (u ^{0} , \vec{u} _{\perp} , u _{z})$. From the equation of motion (\ref{EQLV}) we find that the Green's function satisfies
\begin{align}
\left[ \partial ^{2} _{t} - \partial ^{2} _{z} - \vec{\nabla} ^{2} _{\perp} + \lambda (u _{0} \partial _{t} - \vec{u} _{\perp} \cdot \vec{\nabla} _{\perp} - u _{z} \partial _{z} ) ^{2} + m ^{2} \right] G (x,x^{\prime}) = \delta (x - x ^{\prime}) , \label{EQG2a}
\end{align}
where $\vec{\nabla} _{\perp} ^{2} = \vec{e} _{x} \partial _{x} ^{2} + \vec{e} _{y} \partial _{y} ^{2}$ is the transverse Laplacian.

The symmetry of the system suggests that the GF must possess translational invariance in the transverse $x$ and $y$ directions. Exploiting this symmetry we further introduce the reduced Green's function $g(z,z ^{\prime})$ according to the Fourier transform \cite{Schwinger}
\begin{equation}
G(x,x ^{\prime})=\int\frac{d^2 \vec{k}_\bot}{(2\pi)^2}  e^{i \vec{k} _{\perp} \cdot (\vec{r} - \vec{r} ^{\, \prime}) _{\perp}}\int \frac{d\omega}{2\pi}e^{-i\omega(t-t^{\prime})}g(z,z^{\prime}) , \label{GFGeneral}
\end{equation}
where we have suppressed the dependence of $g$ on the frequency $\omega$ and the transverse momentum $\vec{k} _{\perp}$ for the sake of brevity. The notation is $\vec{f} _{\perp} = f _{x} \vec{e} _{x} + f _{y} \vec{e} _{y}$ for any vector $\vec{f}$ (i.e. $\vec{f} _{\perp}$ is the transverse part of $\vec{f}\,$). Now we have to determine the reduced GF $g(z,z ^{\prime})$.  The substitution of Eq. (\ref{GFGeneral}) into Eq. (\ref{EQG2a}) yields the reduced GF equation:
\begin{align}
\left[ \gamma ^{2} + (1 - \lambda u _{z} ^{2}) \partial ^{2} _{z} - 2 i  \lambda u _{z} ( \omega u _{0} + \vec{k} _{\perp} \cdot \vec{u} _{\perp} ) \partial _{z} \right] g (z,z^{\prime}) = - \delta (z - z ^{\prime}) , \label{reduGF}
\end{align}
where $\gamma ^{2} = \omega ^{2} - k ^{2} _{\perp}  - m ^{2} + \lambda ( \omega u _{0} + \vec{k} _{\perp} \cdot \vec{u} _{\perp} ) ^{2}$. This equation is to be solved subject to the appropriate boundary conditions (e.g. on the plates, at infinity, etc). To this end, we follow the usual discontinuity method, which consists in solving the differential equation (\ref{reduGF}) on the line without the singular point $z = z ^{\prime}$, and then matching the solutions with the appropriate boundary conditions there. Indeed, if one accepts that $g$ is bounded when $z$ is in the infinitesimal neighborhood of $z ^{\prime}$, integration of (\ref{reduGF}) over the interval $z ^{\prime} - 0 ^{+}$ and $z ^{\prime} + 0 ^{+}$ yields 
\begin{align}
 - \frac{\partial g (z,z ^{\prime})}{\partial z} \Bigg| _{z = z ^{\prime} - 0^{+}} ^{z = z ^{\prime} + 0^{+}} &= \frac{1}{1 - \lambda u _{z} ^{2}} . 
\end{align}
Then the continuity of $g$ at $z = z ^{\prime}$ follows. The two independent solutions of the differential equation (\ref{reduGF}) in the region $z \neq z ^{\prime}$ are given by $e ^{i \xi _{0} z} e ^{\pm i \xi _{1} z}$, where 
\begin{align}
\xi _{0} = \frac{\lambda u _{z} ( \omega u _{0} + \vec{k} _{\perp} \cdot \vec{u} _{\perp} )}{1 - \lambda u _{z} ^{2}} , \quad \xi _{1} = \sqrt{ \frac{\gamma ^{2} + (1 - \lambda u _{z} ^{2}) \xi _{0} ^{2}}{1 - \lambda u _{z} ^{2}} } . 
\end{align}
Therefore, the solution to (\ref{reduGF}) can be expressed in terms of the solutions, $e ^{i ( \xi _{0} + \xi _{1}) z}$ and $e ^{i ( \xi _{0} - \xi _{1}) z}$. With the above results, we are ready to compute the needed different Green's functions.

First, we consider the case of the reduced GF between two parallel conductive plates. So, we have to solve Eq. (\ref{reduGF}) subject to the Dirichlet boundary conditions on the plates, i.e. $g _{\parallel} (0,z ^{\prime}) = g _{\parallel} (L,z ^{\prime}) = 0$, where the subscript $\parallel$ indicates that this corresponds to the GF for the parallel plates configuration. In this way, the reduced GF between the plates can be written as
\begin{align}
g _{\parallel} (z,z ^{\prime}) = e ^{i \xi _{0} z} \left\lbrace \begin{array}{l} A e ^{i \xi _{1} z} + B e ^{- i \xi _{1} z} \\[6pt] C e ^{i \xi _{1} z} + D e ^{- i \xi _{1} z} \end{array} \right. \;\; \begin{array}{l} 0<z< z^{\prime} \\[6pt] z ^{\prime} <z<L \end{array} . 
\end{align} 
where the coefficients $A$, $B$, $C$ and $D$, are to be determined by imposing the four boundary conditions. After some algebra we obtain
\begin{align}
g _{\parallel} (z,z ^{\prime}) &= - e ^{i \xi _{0} (z - z ^{\prime})} \frac{\sin ( \xi _{1} z _{<}) \sin [ \xi _{1} (z _{>} -L) ]}{(1 - \lambda u _{z} ^{2}) \xi _{1} \sin (\xi _{1} L)}, \label{gz1}
\end{align}
where $z _{>}$ ($z _{<}$) is the greater (lesser) between $z$ and $z ^{\prime}$. The result for Neumann boundary conditions on the plates is obtained by exchanging the functions $\sin x$ for $\cos x $ in the numerator of Eq. (\ref{gz1}). One can further verify that, in the limit $\lambda \rightarrow 0$, the reduced GF (\ref{gz1}) correctly reduces to the Lorentz invariant case
\begin{align}
g _{0} (z,z ^{\prime}) &= - \frac{\sin ( \beta z _{<}) \sin [ \beta (z _{>} -L) ]}{\beta \sin (\beta L) } , \label{gz0}
\end{align}
where $\beta ^{2} =  \omega ^{2} - k ^{2} _{\perp}- m ^{2}$ \cite{Schwinger}. As we can see, the reduced GF $g _{\parallel} (z,z ^{\prime})$ cannot be expressed in terms of $g _{0} (z,z ^{\prime})$ due to the intricate dependence of the former on $\lambda$ and $u ^{\mu}$. However, as we will see later, $G _{\parallel} (x,x ^{\prime})$ can be expressed in terms of $G _{0} (x,x ^{\prime})$ in the limit $x ^{\prime} \rightarrow x$ after an appropriate change of variables in the frequency and transverse momentum. Here, $G _{\parallel}$ and $G _{0}$ are the GFs in coordinate representation, as defined in Eq. (\ref{GFGeneral}).

In order to evaluate renormalized physical quantities, such as the Casimir energy and the Casimir stress, we will also need the Green's functions in free space (i.e. in the absence of the plates) and that in the presence of one plate, respectively. 
A calculation just like that which led to Eq. (\ref{gz1}) yields, in vacuum, to
\begin{equation}
    g _{\mbox{\scriptsize v}} (z,z^\prime ) = \frac{i}{2\xi_1} \frac{e^{i\xi_0(z-z^\prime)}}{(1-\lambda u_z^2)} e^{i\xi_1(z _{>}- z _{<})}, \label{freespace}
\end{equation}
which has the correct outgoing boundary conditions as $z \to \pm \infty$, i.e. $e ^{i (\xi _{0} \pm \xi _{1}) z}$. Similarly, we can compute the Green's function which vanishes at $z=L$, and has outgoing boundary condition as $z\rightarrow\infty$, i.e. $g _{\vert} \sim e^{i (\xi _{0} + \xi _{1}) z}$. The result for $z,z ^{\prime} > L$ is
\begin{equation}
    g _{\vert} (z,z^\prime ) = \frac{1}{\xi_1} \frac{e^{i\xi_0(z-z^\prime)}}{(1-\lambda u_z^2)} \sin [ \xi_1(z _{<}-L) ] e^{i\xi_1(z _{>}- L)} .  \label{1plate}
\end{equation}
As before, $z _{>}$ ($z _{<}$) is the greater (lesser) between $z$ and $z ^{\prime}$.

\section{Vacuum Stress-Energy Tensor} \label{VSsection}

In Sec. \ref{Intro} we presented the stress-energy tensor for this theory. Now we address its vacuum expectation value (vev), which from now on we will refer as the vacuum stress (VS).

The local approach to compute the VS was initiated in Ref. \cite{BrownMaclay}, where the authors calculated the renormalized stress-energy tensor by means of GF techniques. They used the fact that the Green's function is related to the vacuum expectation value of the time-ordered product of fields according to
\begin{equation}
G (x,x') = -i \langle 0 |\hat{\mathcal{T}} \phi (x) \phi ( x ^{\prime}) |0\rangle .
\end{equation}  
Therefore the VS can be obtained from appropriate derivatives of the GF. Using the standard point splitting technique and taking the vacuum expectation value of the stress-energy tensor (\ref{Tmunu}) we find
\begin{align}
    \langle T ^{\mu\nu} \rangle = -i \lim _{x ^{\prime} \rightarrow x} [\partial ^{\mu}  \partial^{\prime \nu} + \lambda u ^{\mu} \partial ^{\nu}  \left( u \cdot \partial ^{\prime} \right) ]  G(x, x^\prime) - \eta ^{\mu \nu} \langle \mathcal{L} \rangle , \label{VEVTmunu}
\end{align}
where
\begin{align}
    \langle \mathcal{L} \rangle = - i \lim _{x ^{\prime} \to x}  \frac{1}{2} \left[ \partial \cdot \partial ^{\prime} + \lambda \left(u \cdot \partial \right) \left(u \cdot \partial ^{\prime} \right) - m ^{2} \right] G (x , x ^{\prime}) .
\end{align}
In this context, the energy density (energy per unit volume) is defined as the time-time component of the VS, i.e. $\langle T ^{00} \rangle$. From the above expressions, together with the $3+1$ representation of the Green's function (\ref{GFGeneral}), we obtain the following general expression for the energy density
\begin{align}
\langle T^{00} \rangle &= - i \lim _{z ^{\prime} \to z} \int \frac{d \omega}{2 \pi} \int \frac{d ^{2} \vec{k} _{\perp}}{(2 \pi) ^{2}} \left[ \omega ^{2} + \lambda u _{0} \omega  ( u _{0} \omega + \vec{u} _{\perp} \cdot \vec{k} _{\perp} ) - i \lambda u _{0} \omega  u _{z}  \partial _{z} \right] g (z,z ^{\prime}) - \langle \mathcal{L} \rangle , \label{T00a}
\end{align}
where the vev of the Lagrangian is
\begin{align}
\langle \mathcal{L} \rangle &= - \frac{i}{2} \lim _{z ^{\prime} \to z} \int \frac{d \omega}{2 \pi} \int \frac{d ^{2} \vec{k} _{\perp}}{(2 \pi) ^{2}} \left[ \gamma ^{2} - (1 - \lambda u _{z} ^{2} ) \partial _{z} \partial _{z ^{\prime}} + i \lambda u _{z} ( u _{0} \omega + \vec{u} _{\perp} \cdot \vec{k} _{\perp} ) \left( \partial _{z ^{\prime}} - \partial _{z} \right)   \right] g (z , z ^{\prime} ) . \label{ExpValL}
\end{align}
Also, by virtue of the boundary conditions on bounding surfaces, the pressure (force per unit area) on the boundary can be obtained from the normal-normal component of the VS, i.e. $\langle n _{\mu} n _{\nu} T ^{\mu \nu} \rangle$ being $n _{\mu}$ the unit normal to the surface. Taking $n ^{\mu} = (0,0,0,1)$, an explicit expression for the pressure is
\begin{align}
\langle T^{zz} \rangle &= - i \lim _{z ^{\prime} \to z} \int \frac{d \omega}{2 \pi} \int \frac{d ^{2} \vec{k} _{\perp}}{(2 \pi) ^{2}}  \left[ (1 - \lambda u _{z} ^{2}) \partial _{z} \partial _{z ^{\prime}} + i \lambda u _{z} (u _{0} \omega + \vec{u _{\perp}} \cdot \vec{k} _{\perp}  ) \partial _{z} \right] g (z,z ^{\prime}) + \langle \mathcal{L} \rangle , \label{ExpValTzz}
\end{align}
where $\langle \mathcal{L} \rangle$ is given in Eq. (\ref{ExpValL}). 
These results will be extensively used in the next section to evaluate the Casimir stress upon the plates by i) variation of the energy density (\ref{T00a}) and ii) direct evaluation of the normal-normal component of the VS (\ref{ExpValTzz}). We will also analize the local effects in the energy density. Interestingly, as we will see, the vacuum expectation value of the Lagrangian (\ref{ExpValL}) is the responsible of divergences of the energy density near the boundaries, which of course are not physical.

%{\color{red} It is worth to mention that all the LV contributions emerge from two sides: a) the terms proportional to $\lambda$ in the VS in Eq. (\ref{TmunuExp}), including those in $\langle\mathcal{L}\rangle$  and b) the reduced GF in Eq. (\ref{gz1}).}

\section{Casimir Effect} \label{CEsection}

In its most basic form, the Casimir effect is the attraction between two neutral perfectly conductive parallel plates placed in vacuum \cite{Casimir}. The attractive force can be considered as arising due to the change in the zero-point energy of the electromagnetic field when the plates are brought into position.

There are different ways in which the Casimir energy may be computed. The most commonly used is perhaps the mode-summation method, which consist in the direct evaluation of infinite sums over eigenvalues of zero-point field modes. The local approach, which is based upon the use of Green's functions, represents a formally elegant manner to derive the Casimir energy and Casimir stress \cite{Milton}. Both treatments can be shown to be formally equivalent; however, local methods are richer than global ones since they provide much more information about the system. Using the results derived above, in this section we compute the Casimir stress upon the plates when a Lorentz-violating scalar field is confined between them.

\subsection{Global Casimir energy} \label{GlobalCasSec}

Following Weiskopf, Schwinger and others \cite{Schwinger2, Schwinger3, Weisskopf, Heisenberg}, the physical vacuum energy is defined as the difference between the zero-point energy in the presence of boundaries and that of the free vacuum. In the language of quantum field theory, the Casimir energy stored in the field between the plates is expressed as
\begin{equation}
    \mathcal{E} _{C} (L) = \int _{0} ^{L} \langle T ^{00} \rangle _{\textrm{ren}} \, dz ,
    \label{Eglobal}
\end{equation}
where 
\begin{equation}
    \langle T^{00} \rangle _{\textrm{ren}} = \langle T^{00} \rangle _{\parallel} - \langle T^{00} \rangle _{\textrm{v}}
\end{equation}
is the renormalized time-time component of the VS, which is just the difference between the energy density in the presence of the plates $\langle T^{00} \rangle _{\parallel}$ and that of the free vacuum $\langle T^{00} \rangle _{\textrm{v}}$ \cite{Milton}. This means that the former must be computed by using the Green's function for the parallel plates configuration, given by Eq. (\ref{gz1}), while the later must me computed with the vacuum GF (\ref{freespace}). Let us start with $\langle T^{00} \rangle _{\parallel}$. Substituting the reduced Green's function in the presence of the plates $g _{\parallel} (z,z ^{\prime})$ into Eq. (\ref{T00a}) we obtain
\begin{align}
\langle T^{00} \rangle_{\parallel} &= - i \int \frac{d \omega}{2 \pi} \int \frac{d ^{2} \vec{k} _{\perp}}{(2 \pi) ^{2}} \left[ \omega ^{2} + \frac{\lambda u _{0}}{1 - \lambda u _{z} ^{2}} \omega  ( u _{0} \omega + \vec{u} _{\perp} \cdot \vec{k} _{\perp} )  \right] g _{\parallel} (z,z) \notag \\ & \phantom{==} + \frac{ \lambda u _{0} u _{z}}{1 - \lambda u _{z} ^{2}} \int \frac{d \omega}{2 \pi} \int \frac{d ^{2} \vec{k} _{\perp}}{(2 \pi) ^{2}} \, \omega \, \frac{\sin ( \xi _{1} z ) \cos [ \xi _{1} (z -L) ]}{\sin (\xi _{1} L)} - \langle \mathcal{L} \rangle _{\parallel} . \label{T00Plates}
\end{align}
and from Eq. (\ref{ExpValL}) we find the vev of the Lagrangian
\begin{align}
\langle \mathcal{L} \rangle _{\parallel} &= - i \int \frac{d \omega}{2 \pi} \int \frac{d ^{2} \vec{k} _{\perp}}{(2 \pi) ^{2}} \frac{\xi _{1}}{2} \frac{\cos [ \xi _{1} ( 2z - L) ]}{\sin (\xi _{1} L)} .  \label{LagPlates}
\end{align}
Two of the three terms appearing in Eq. (\ref{T00Plates}) will not contribute to the Casimir stress upon the plates, as we shall discuss just now. On the one hand, one can easily check that the integral of Eq. (\ref{LagPlates}) in the interval $[0,L]$, as required by Eq. (\ref{Eglobal}), produces $\int _{0} ^{L} \langle \mathcal{L} \rangle _{\parallel} \, dz = (1/2i) \int \frac{d \omega}{2 \pi} \int \frac{d ^{2} \vec{k} _{\perp}}{(2 \pi) ^{2}}$, which is a formally divergent term which nevertheless does not depend on $L$, and as such it will not contribute to the Casimir pressure. On the other hand, the second integral (whose integrand is proportional to $\omega$) yields zero by symmetry considerations. This is clarify below. Therefore, these two terms can be safely disregarded in Eq. (\ref{T00Plates}), and hence we will focus only in the first term.

In order to develop the integral in Eq. (\ref{T00Plates}), let us define the rescaled quantities $\tilde{z} = z / \sqrt{1 - \lambda u _{z} ^{2}}$ and $\tilde{L} = L / \sqrt{1 - \lambda u _{z} ^{2}}$, such that the Green's function $g _{\parallel}$ (given by Eq. (\ref{gz1})) at coincident arguments can be written in terms of the Lorentz-symmetric GF $g _{0}$ (given by Eq. (\ref{gz0})) at rescaled coincident arguments as follows
\begin{align}
    g _{\parallel} (z,z) = \frac{1}{\sqrt{1 - \lambda u _{z} ^{2}}} \tilde{g} _{0} (\tilde{z},\tilde{z}) ,
\end{align}
where
\begin{align}
    \tilde{g} _{0} (\tilde{z},\tilde{z}) = - \frac{\sin ( \alpha \tilde{z}) \sin [ \alpha (\tilde{z} - \tilde{L}) ]}{\alpha \sin (\alpha \tilde{L}) } \label{g0Tilde}
\end{align}
with however $\alpha ^{2} = \omega ^{2} - k _{\perp} ^{2} - m ^{2} + \frac{\lambda}{1 - \lambda u _{z} ^{2}} (\omega u _{0} + \vec{k} _{\perp} \cdot \vec{u} _{\perp}) ^{2}$. Therefore, the main difference between $\tilde{g} _{0} (\tilde{z},\tilde{z})$ and the Lorentz-symmetric GF $g _{0}(z,z)$ is that the $\alpha$ has a highly nontrivial dependence on the frequency $\omega$ and transverse momentum $\vec{k} _{\perp}$ as compared with the $\beta ^{2} = \omega ^{2} - k _{\perp} ^{2} - m ^{2}$ appearing in the Lorentz invariant GF. Indeed, one can confirm that in the limit $\lambda \to 0$, $\alpha \to \beta$ and hence $\tilde{g} _{0} (\tilde{z},\tilde{z})$ correctly reduces to $g _{0} (z,z)$, as it should be.

To proceed further we define the three-vector $\vec{\kappa} = (\omega , k _{x} , k _{y}) \in \mathbb{R} ^{3}$, such that the integral in Eq. (\ref{T00Plates}) can be written as 
\begin{align}
\langle T^{00} \rangle_{\parallel} &= - i \int \frac{d \omega}{2 \pi} \int \frac{d ^{2} \vec{k} _{\perp}}{(2 \pi) ^{2}} \Delta _{ij} \kappa _{i} \kappa _{j} \, \frac{\tilde{g} _{0} (\tilde{z},\tilde{z})}{\sqrt{1 - \lambda u _{z} ^{2}}} , \label{T00Plates2}
\end{align}
where $\boldsymbol{\Delta} = (\Delta _{ij})$ is a real-valued 3$\times$3 symmetric matrix whose explicit form is read-off directly from Eq. (\ref{T00Plates}). Also, the real ternary quadratic form $\alpha$ can be written as $\alpha ^{2} = \Lambda _{ij} \kappa _{i} \kappa _{j} - m ^{2}$, where $\boldsymbol{\Lambda} = (\Lambda _{ij})$ is another 3$\times$3 real-valued symmetric matrix which we easily read from the definition of $\alpha$. The main difficulty to evaluate this integral is that the quadratic form $\alpha$, which appears within the $\tilde{g} _{0}$ function, contains crossed-terms (since $\boldsymbol{\Lambda}$ is nondiagonal). In order to solve this problem we use the Jacobi's theorem, which asserts that every quadratic form in $n$ variables has an orthogonal diagonalization \cite{Meara}. This is accomplished with a change of variables $\kappa _{i} = \Gamma _{ij} \kappa _{j} ^{\prime}$, defined by an orthogonal matrix ${\boldsymbol{\Gamma}}= ( \Gamma _{ij})$ (which is technically constructed with the normalized eigenvectors of $\boldsymbol{\Lambda}$). In the problem at hand the corresponding matrix is
\begin{align}
    \boldsymbol{\Gamma} = \left( \begin{array}{ccc} \frac{2 + \lambda ( u ^{2} - u _{z}^2) - \delta}{\varepsilon _{+}} & \frac{ 2 + \lambda ( u ^{2} - u _{z} ^2) +  \delta}{\varepsilon _{-}} & 0 \\  \frac{2 \lambda u _{0} u _{x}}{\varepsilon _{+}} &  \frac{2 \lambda u _{0} u _{x}}{\varepsilon _{-}} & - \frac{u _{y}}{u _{\perp}} \\  \frac{2 \lambda u _{0} u _{y}}{\varepsilon _{+}} & \frac{2 \lambda u _{0} u _{y}}{\varepsilon _{-}} & \frac{u _{x}}{u _{\perp}}
\end{array} \right) ,
\end{align}
where $u ^{2} = u _{\mu} u ^{\mu}$ and
\begin{align}
    \delta ^{2} &= 4 + \lambda \left\lbrace 4 (u ^{2} - u _{z} ^{2}) + \lambda [2 u _{0} ^{2} - (u - u _{z}) ^{2}] [2 u _{0} ^{2} - (u + u _{z}) ^{2}] \right\rbrace , \notag \\[4pt] \varepsilon _{\pm} ^{2} &= [ \delta \pm \lambda ( u _{z} ^{2} - u ^{2}) \mp 2 ] ^{2} + ( 2 \lambda u _{0} u _{\perp} ) ^{2} .
\end{align}
So, in the primed coordinate system the quadratic form $\alpha ^{2}$ takes the simple diagonal form $\alpha ^{2} = \Xi _{ij} \kappa _{i} ^{\prime} \kappa _{j} ^{\prime} - m ^{2} = \Xi _{1} \omega ^{\prime \, 2} + \Xi _{2} k _{x} ^{\prime \, 2} + \Xi _{3} k _{y} ^{\prime \, 2} - m ^{2}$, where we have used that $\Xi _{ij} = \Gamma _{in} \Lambda _{nm} \Gamma _{mj} = \mbox{diag}(\Xi _{1} , \Xi _{2} , \Xi _{3})$, with
\begin{align}
    \Xi _{1} = \frac{\lambda (u _{0} ^{2} + u _{\perp} ^{2}) - \delta}{2 (1 - \lambda u _{z} ^{2})} , \qquad \Xi _{2} & = \frac{\lambda (u _{0} ^{2} + u _{\perp} ^{2}) + \delta}{2 (1 - \lambda u _{z} ^{2})} , \qquad \Xi _{3} = - 1 .
\end{align}
Since the Jacobian of the transformation $J = \det\boldsymbol{\Gamma}$ is $1$, the integral (\ref{T00Plates2}) then becomes
\begin{align}
\langle T^{00} \rangle_{\parallel} &= - i \int \frac{d \omega ^{\prime}}{2 \pi} \int \frac{d ^{2} \vec{k} _{\perp} ^{\, \prime}}{(2 \pi) ^{2}} \, \Pi _{ij} \kappa _{i} ^{\prime} \kappa _{j} ^{\prime} \, \frac{\tilde{g} _{0} (\tilde{z},\tilde{z})}{\sqrt{1 - \lambda u _{z} ^{2}}} , \label{T00Plates3}
\end{align}
where $\Pi _{ij} = \Delta _{nm} \Gamma _{ni} \Gamma _{mj}$. Also, considering that $\tilde{g} _{0}$ is even under the change $\kappa _{i} ^{\prime} \to - \kappa _{i} ^{\prime}$, only even terms in the integrand will survive (i.e. $\Pi _{11} \omega ^{\prime \, 2} + \Pi _{22} k _{x} ^{\prime \, 2} + \Pi _{33} k _{y} ^{\prime \, 2}$), while crossed-terms integrates out to zero (e.g. $\Pi _{12} \omega k _{x}$ and $\Pi _{13} \omega k _{y}$). This analysis justifies the vanishing of the second integral in Eq. (\ref{T00Plates}). Now with a convenient change of variables we can express the function $\tilde{g} _{0}$ exactly as the reduced GF $g _{0}$ in the absence of Lorentz violation, as we shall see. Let us introduce double primed coordinates through the change of variables $\kappa _{i} ^{\prime \prime} = \Omega _{ij} \kappa _{j} ^{\prime}$, where $\Omega _{ij} = \mbox{diag}(\sqrt{\Xi _{1}} , \sqrt{- \Xi _{2}} , \sqrt{- \Xi _{3}})$. So, in the double primed coordinates the quadratic form $\alpha ^{2}$ becomes $\alpha ^{2} = \omega ^{\prime \prime \, 2} - k _{x} ^{\prime \prime \, 2} - k _{y} ^{\prime \prime \, 2} - m ^{2}$, which is exactly equal to the $\beta ^{2}$ appearing in the Lorentz-symmetric case. Therefore, the integral in Eq. (\ref{T00Plates3}) can be written as an integral in terms of the reduced GF $g _{0} (\tilde{z} , \tilde{z})$ evaluated at the rescaled length $\tilde{L}$ as follows
\begin{align}
\langle T^{00} \rangle_{\parallel} &= \frac{- i}{\sqrt{\det \boldsymbol{\Xi}} \sqrt{1 - \lambda u _{z} ^{2}}} \int \frac{d \omega}{2 \pi} \int \frac{d ^{2} \vec{k} _{\perp}}{(2 \pi) ^{2}} \, \left( \frac{\Pi _{11}}{\Xi _{1}} \omega ^{2} - \frac{\Pi _{22}}{\Xi _{2}} k _{x} ^{2} - \frac{\Pi _{33}}{\Xi _{3}} k _{y} ^{2} \right) \, g _{0} (\tilde{z},\tilde{z}) \vert _{\tilde{L}} , \label{T00Plates32}
\end{align}
where we have dropped the double primes since they are integration variables. After performing a Wick's rotation, $\omega \to i \zeta$, it is clear that the integrals over $\zeta ^{2}$, $k _{x} ^{2}$ and $k _{y} ^{2}$ have the same contribution. So, substituting the required matrix elements we obtain
\begin{align}
\langle T^{00} \rangle_{\parallel} &= \frac{1}{\sqrt{1 + \lambda u ^{2}}} \int \frac{d \zeta}{2 \pi} \int \frac{d ^{2} \vec{k} _{\perp}}{(2 \pi) ^{2}} \, \zeta ^{2} \, \frac{\sinh ( \gamma \tilde{z} ) \sinh [ \gamma (\tilde{z}-\tilde{L}) ]}{\gamma \sinh (\gamma \tilde{L}) } , \label{T00Plates4}
\end{align}
where $\gamma ^{2} = \zeta ^{2} + k _{\perp} ^{2} + m ^{2}$. To obtain this result we have used that $(\Pi _{11} / \Xi _{1}) + (\Pi _{22} / \Xi _{2}) + (\Pi _{33} / \Xi _{3}) = 1$ and $(1 - \lambda u _{z} ^{2}) \det \boldsymbol{\Xi} = 1 + \lambda u ^{2}$.

\begin{figure}[h!]
    \centering
    \includegraphics[scale=0.4]{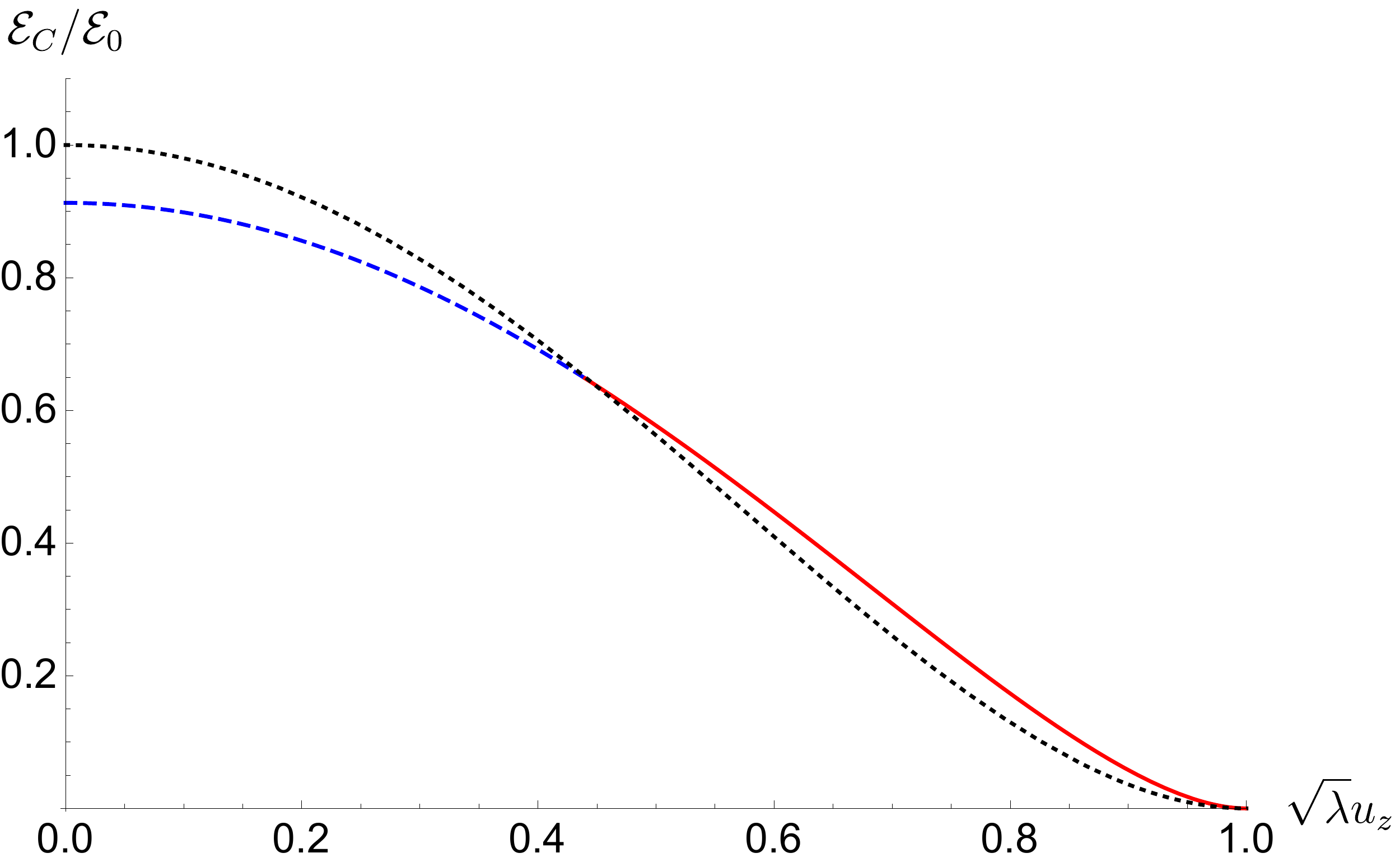}
    \caption{Casmir energy for a massless Lorentz-violating scalar field (in units of the Lorentz-symmetric CE $\mathcal{E} _{0}$) as a function of $\sqrt{\lambda} u _{z}$ for an arbitrary fixed length. For the timelike (blue-dashed line) and the spacelike (red-continuous line) cases we take $\lambda (u _{0} ^{2} - u _{\perp} ^{2}) = 0.2$. The black-dotted line shows the lightlike case.}
    \label{EnergyMassless}
\end{figure}

Following the same procedure, we can evaluate the vacuum energy density $\langle T^{00} \rangle _{\textrm{v}}$. Substituting the reduced vacuum GF (\ref{freespace}) into Eq. (\ref{T00a}), and performing the above analysis to the integral we find
\begin{align}
\langle T^{00} \rangle _{\textrm{v}} &= - \frac{1}{\sqrt{1 + \lambda u ^{2}}} \int \frac{d \zeta}{2 \pi} \int \frac{d ^{2} \vec{k} _{\perp}}{(2 \pi) ^{2}} \, \frac{\zeta ^{2}}{2 \gamma} . \label{T00Vacuum}
\end{align}
Finally we substitute the above results (\ref{T00Plates4}) and (\ref{T00Vacuum}) into Eq. (\ref{Eglobal}). Performing the integration over $z$ and neglecting a constant term (independent of $L$) we finally obtain 
\begin{align}
    \mathcal{E} _{C} (L) = - \sqrt{\frac{1 - \lambda u _{z} ^{2}}{1 + \lambda u ^{2}}} \int \frac{d \zeta}{2 \pi} \int \frac{d ^{2} \vec{k} _{\perp}}{(2 \pi) ^{2}} \frac{\zeta ^{2}}{2 \gamma} \tilde{L} \, [ \coth{(\gamma \tilde{L})} - 1 ] .
     \label{EFinal2a}
\end{align}
The resulting integrals can be easily computed; however, we recognize it as the standard Casimir energy density  where Lorentz symmetry is preserved, with the only difference that the distance $L$ between plates has been rescaled by the factor $\sqrt{1 - \lambda u _{z} ^{2}}$. In this way we establish that
\begin{align}
    \mathcal{E} _{C} (L) = \sqrt{\frac{1 - \lambda u _{z} ^{2}}{1 + \lambda u ^{2}}} \mathcal{E} _{0} (\tilde{L}) ,  \label{CasimirEnergyFin}
\end{align}
where $\tilde{L} = L / \sqrt{1 - \lambda u _{z} ^{2}}$ and
\begin{align}
    \mathcal{E} _{0} (L) = \left\lbrace \begin{array}{ll} - \frac{\pi ^{2}}{1440 L ^{3}} & \mbox{if }\mbox{ $m$ $=0$} \\ [6pt] - \frac{m ^{2}}{8 \pi ^{2} L} \sum _{n = 1} ^{\infty} \frac{1}{n ^{2}} K _{2} (2 m n L) & \mbox{if }\mbox{ $m$ $\neq0$} \end{array} \right. ,
\end{align}
being $K _{2} (x)$ the second-order Bessel function of the second kind. As expected, our results reduce to the Lorentz-invariant energy densities in the limit $\lambda \rightarrow 0$ \cite{Milton, Farina}. The summation appearing in the Casimir energy for a massive scalar field do not have closed analytical form; however, it can be approximated in the limit of small and large masses. In the limit $m \tilde{L} \ll 1$ the Casimir energy (\ref{CasimirEnergyFin}) becomes
\begin{align}
    \mathcal{E} _{C} (L) \approx - \sqrt{\frac{1 - \lambda u _{z} ^{2}}{1 + \lambda u ^{2}}} \left( \frac{\pi ^{2}}{1440 \tilde{L} ^{3}} - \frac{m ^{2}}{96 \tilde{L}} \right)  ,
\end{align}
while in the limit $m \tilde{L} \gg 1$ we find
\begin{align}
    \mathcal{E} _{C} (L) \approx - \sqrt{\frac{1 - \lambda u _{z} ^{2}}{1 + \lambda u ^{2}}} \frac{m ^{2}}{16 \pi ^{2} \tilde{L}} \sqrt{\frac{\pi}{m \tilde{L}}} e ^{- 2m \tilde{L}} .
\end{align}
Now let us analyze our results. We observe that in the massless case, the Casimir energy always leads to attraction (negative pressure). In fig. \ref{EnergyMassless}, by fixing the value of $\lambda (u _{0} ^{2} - u _{\perp} ^{2}) > 0$, we plot the ratio $\mathcal{E} _{C} (L) / \mathcal{E} _{0} (L)$ as a function of $\sqrt{\lambda} u _{z} \in [0,1]$ for timelike (blue-dashed line), spacelike (red-continuous line) and lightlike (black-dotted line) cases. We observe that, in the timelike case $u ^{2} > 0$, the ratio decreases monotonically from $r _{0} = \frac{1}{\sqrt{1 + \lambda (u _{0} ^{2} - u _{\perp} ^{2})}} < 1$, at $u _{z} = 0$, to $r _{c} = 1 - \lambda (u _{0} ^{2} - u _{\perp} ^{2})$ at $ u _{z} = \sqrt{u _{0} ^{2} - u _{\perp} ^{2}}$ (or $u ^{2} = 0$). There, the sign of $u ^{2}$ flips and the spacelike case $u ^{2} <0$ starts. In this case, the ratio decreases monotonically from $r _{c}$ to zero at $\sqrt{\lambda} u _{z} = 1$. The lightlike case decreases from 1, at $\sqrt{\lambda} u _{z} = 0$, to zero at $\sqrt{\lambda} u _{z} = 1$. From Eq. (\ref{CasimirEnergyFin}) we observe that in a lightlike particular case, for which $u _{z} = 0$ and $u _{0} ^{2} - u _{\perp} ^{2} = 0$, the Casimir energy (and hence the pressure) does not see Lorentz violation, i.e. $\mathcal{E} _{C} (L) = \mathcal{E} _{0} (L)$. So, in all cases, the Casimir energy in the presence of Lorentz violation is always smaller than the Lorentz-symmetric case.

\subsection{Stress on the plates} \label{ST}
 
 Now let us derive the Casimir stress upon the plate at $z = L$ by direct evaluation of the normal-normal component of the stress-energy tensor, whose general expression is given by Eq. (\ref{ExpValTzz}). To this end, we have to compute the discontinuity of $\langle T ^{zz} \rangle$ at that plate. Let $\langle T ^{zz} \rangle _{\parallel}$ be the vacuum stress due to the confined scalar field, and be $\langle T ^{zz} \rangle _{\vert}$ the vacuum stress due to the scalar field at the right side of the plate \cite{Milton, Farina}. So the Casimir stress upon the plate at $z = L$ is
 \begin{align}
    \mathcal{F} _{C} (L) = \langle T ^{zz} \rangle_{\parallel} - \langle T ^{zz} \rangle _{\vert} . \label{CasimirStress}
\end{align} 
Let us calculate each one separately. The stress upon the plate due to the confined field must be computed substituting the reduced Green's function $g _{\parallel} (z,z ^{\prime})$ into Eq. (\ref{ExpValTzz}) and evaluating the result at $z = L$. To this end, we follow the same procedure as that in the previous section. A straightforward calculation yields
\begin{align}
\langle T^{zz} \rangle_{\parallel} &= - \frac{1}{\sqrt{1 + \lambda u ^{2}}} \int \frac{d \zeta}{2 \pi} \int \frac{d ^{2} \vec{k} _{\perp}}{(2 \pi) ^{2}}  \frac{\gamma}{2}  \coth (\gamma \tilde{L}) .
\end{align}
Of course, this integral diverges. The proper subtraction of the pressure due to the field outside yields a finite value. Substituting the reduced Green's function $g _{\vert} (z,z ^{\prime})$ into Eq. (\ref{ExpValTzz}) we find
\begin{align}
\langle T^{zz} \rangle_{\vert} &= - \frac{1}{\sqrt{1 + \lambda u ^{2}}} \int \frac{d \zeta}{2 \pi} \int \frac{d ^{2} \vec{k} _{\perp}}{(2 \pi) ^{2}}  \frac{\gamma}{2} . 
\end{align}
Inserting these results into Eq. (\ref{CasimirStress}) we obtain that the Casimir stress upon the plate at $z = L$ is
\begin{align}
    \mathcal{F} _{C} (L) = - \frac{1}{\sqrt{1 + \lambda u ^{2}}} \int \frac{d \zeta}{2 \pi} \int \frac{d ^{2} \vec{k} _{\perp}}{(2 \pi) ^{2}}  \frac{\gamma}{2} [ \coth (\gamma \tilde{L}) - 1 ] . \label{IntegralForce}
\end{align}
The resulting integral corresponds to the pressure in the absence of Lorentz violation; however, evaluated at the rescaled length $\tilde{L}$. The final expression for the stress is then
\begin{align}
    \mathcal{F} _{C} (L) = \frac{1}{\sqrt{1 + \lambda u ^{2}}} \mathcal{F} _{0} (\tilde{L}) , \label{CasimirForceFin}
\end{align}
where
\begin{align}
    \mathcal{F} _{0} (L) = \left\lbrace \begin{array}{ll} - \frac{\pi ^{2}}{480 L ^{4}} & \mbox{if }\mbox{ $m$ $=0$} \\[6pt] \frac{1}{4 \pi ^{2}} \int _{0} ^{\infty} \frac{\tau ^{2} \sqrt{\tau ^{2} + m ^{2}}}{e ^{2 L \sqrt{\tau ^{2} + m ^{2}}} - 1} d \tau  & \mbox{if }\mbox{ $m$ $\neq0$} \end{array}  \right. .
\end{align}
One can further see that this result coincides with the negative derivative of the Casimir energy (\ref{CasimirEnergyFin}) with respect to $L$, i.e.
\begin{align}
    \mathcal{F} _{C} (L) = - \frac{\partial \mathcal{E}_{C} (L)}{\partial L} .
\end{align}

As a consistency check we verify that our result (\ref{CasimirForceFin}) agrees with the one expected from the coordinate redefinition method in the regime $\lambda u _{z} ^{2} \ll 1$ \cite{Altschul1,Altschul2}. We recall that in this case, the change of spacetime coordinates $x ^{\prime \, \mu} = x ^{\mu} - \frac{1}{2} \lambda u ^{\mu} u _{\nu} x ^{\nu}$ transforms the  Lorentz-violating scalar field theory into the Lorentz-invariant theory. So, in the primed coordinate system the Casimir energy will be the one predicted by the Lorentz-invariant theory, with however a redefinition of the plate-to-plate separation according to the spacetime transformation and an overall factor arising from the Jacobian of the transformation. One can directly verify that the multiplicative factor appearing in Eq. (\ref{CasimirForceFin}) corresponds with the inverse square root of the Jacobian $J = 1 + \lambda u ^{2} $, which is consistent with the definition of the Green's function as the vacuum expectation value of the product of two fields. Also, if the distance between the plates is $L$ in the unprimed coordinate system, the corresponding length in the primed system will be $L ^{\prime} = L \left( 1 + \frac{1}{2} \lambda u _{z} ^{2} \right)$, which coincides with the leading order approximation of our $\tilde{L}$.

As a final remark, we point out that although the expectation value of the Lagrangian (\ref{ExpValL}) does not contribute to the Casimir energy (\ref{CasimirEnergyFin}), it does to the Casimir stress (\ref{CasimirForceFin}). As we shall see in the next section, it also plays a fundamental role regarding the behaviour of the field near the boundaries.

\subsection{Local effects} \label{LE}

In the previous sections we derived an expression for the global Casimir energy by computing the integral of $\langle T^{00} \rangle _{\textrm{ren}}$ in the region between the plates. We have also validated our results by the direct calculation of the Casimir stress. It is worth mentioning that these results can also be obtained by other global methods, for example, summing over the ground state modes or by evaluation of the Lifshitz formula for the Casimir energy. However, Green's function methods allow us to study the local energy density, or, more generally $\langle T^{\mu \nu} \rangle$, which will reveal new information about the divergence structure of the theory \cite{BrownMaclay, Milton, Deutsch&Candelas}. This is precisely the goal of this section.

\begin{figure}[h!]
    \centering
    \includegraphics[scale=0.4]{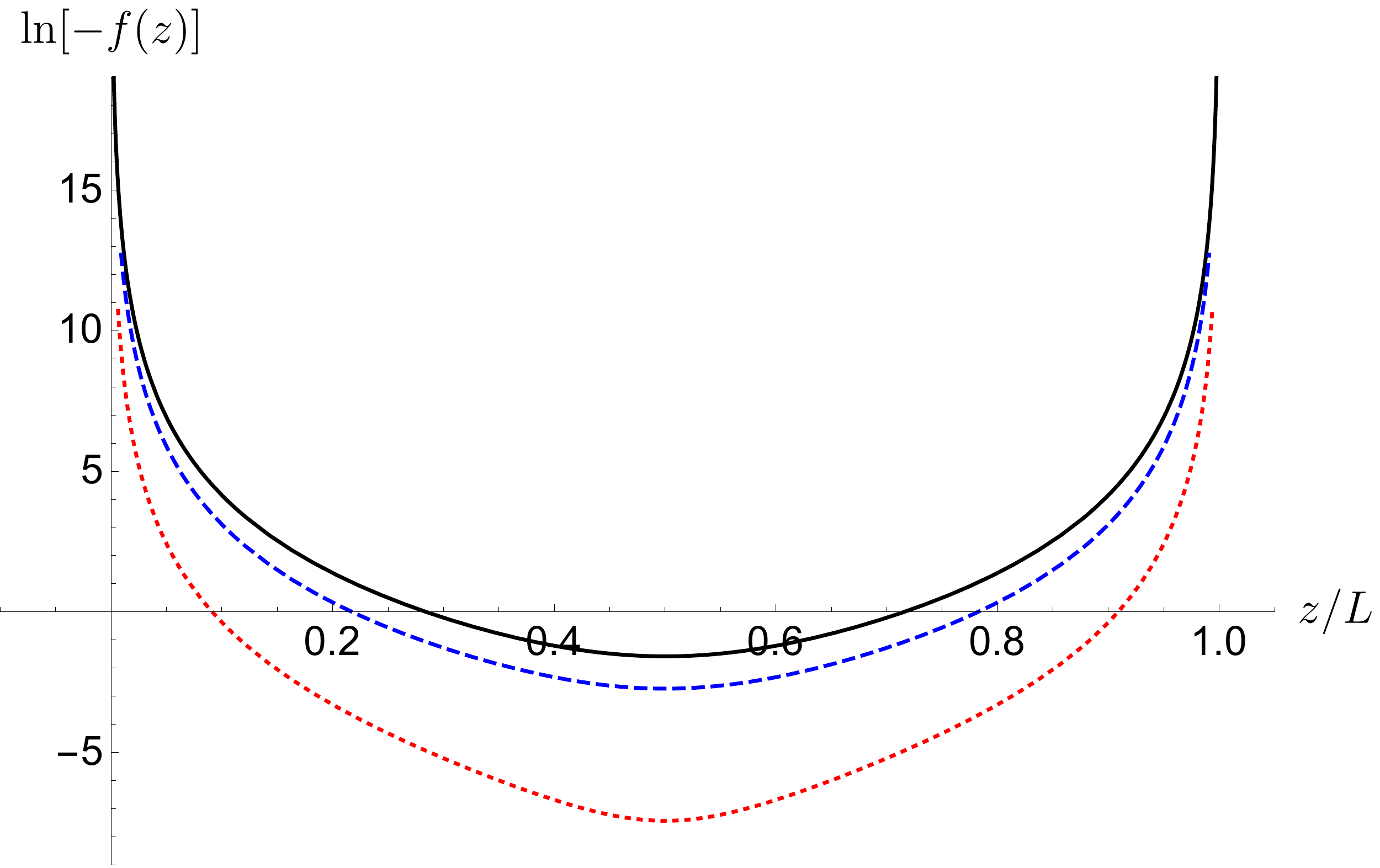}
    \caption{Singular part of the local energy density between the plates. The black-continuous line correspond to the Lorentz-symmetric case. The blue-dashed and red-dotted lines exhibit the effects of Lorentz violation, with $\lambda u _{z} ^{2} = 0.5$ and $\lambda u _{z} ^{2} = 0.95$, respectively.}
    \label{LocalEffectsPlot}
\end{figure}

Let us start with the energy density per unit volume between the plates. Using the results of equations (\ref{LagPlates}) and (\ref{T00Plates4}), after some algebraic simplifications we obtain
\begin{align}
     \langle T^{00} \rangle &= - \frac{1}{\sqrt{1 + \lambda u ^{2}}} \int \frac{d \zeta}{2 \pi} \int \frac{d ^{2} \vec{k} _{\perp}}{(2 \pi) ^{2}} \left\lbrace \frac{\zeta ^{2}}{2 \gamma} \coth (\gamma \tilde{L}) + \frac{k _{\perp} ^{2} + m ^{2}}{2 \gamma} \frac{\cosh [\gamma (2 \tilde{z} - \tilde{L})]}{\sinh (\gamma \tilde{L})} \right\rbrace .
\end{align}
We evaluate this by introducing the polar coordinates $k _{\perp} = \rho \cos \theta$ and $\zeta = \rho \sin \theta$, where $\rho \in [0, \infty)$ and $\theta \in [- \pi / 2 , \pi / 2]$ (since the plane $\zeta k _{\perp}$ covers the right half of $\mathbb{R} ^{2}$). Straightforward calculations yield
\begin{align}
    \langle T^{00} \rangle &= - \frac{1}{12 \pi ^{2}} \frac{1}{\sqrt{1 + \lambda u ^{2}}} \int _{0} ^{\infty} \left\lbrace \frac{\rho ^{4}}{\gamma ^{\ast}} \frac{2}{e ^{2 \gamma ^{\ast} \tilde{L}} - 1} + \frac{\rho ^{4}}{\gamma ^{\ast}} + \frac{\rho ^{2}}{\gamma ^{\ast}} (2 \gamma ^{\ast \, 2} + m ^{2}) \frac{e ^{2 \gamma ^{\ast} \tilde{z}} + e ^{2 \gamma ^{\ast} ( \tilde{L} - \tilde{z} )}}{e ^{2 \gamma ^{\ast} \tilde{L}} - 1} \right\rbrace d \rho ,  \label{LocalEffects}
\end{align}
where $\gamma ^{\ast} = \sqrt{\rho ^{2} + m ^{2}}$. We observe that the second term produces a constant energy density, independent of $L$, so it can be discarded as irrelevant \cite{Milton}. The first term is found to be proportional to the Casimir energy $\mathcal{E} _{C}$. Comparing such term with the expression (\ref{EFinal2a}) for the Casimir energy we find
\begin{align}
    \mathfrak{U} &= - \frac{1}{6 \pi ^{2}} \frac{1}{\sqrt{1 + \lambda u ^{2}}} \int _{0} ^{\infty} \frac{\rho ^{4}}{\gamma ^{\ast}} \frac{1}{e ^{2 \gamma ^{\ast} \tilde{L}} - 1} d \rho = \mathcal{E} _{C} / L .
\end{align}
Also, with a simple change of variables, the third term in Eq. (\ref{LocalEffects}) can be written (and defined) as
\begin{align}
    f (z) &= - \frac{1}{192 \pi ^{2} \tilde{L} ^{4}} \frac{1}{\sqrt{1 + \lambda u ^{2}}} \int _{2 m \tilde{L}} ^{\infty} \sqrt{y ^{2} - (2 m \tilde{L}) ^{2}} \left[ 2y ^{2} + (2 m \tilde{L}) ^{2} \right] \frac{e ^{y z/L} + e ^{y(1-z/L)}}{e ^{y} -1} dy . \label{fz}
\end{align}
All in all, the energy density per unit volume is expressed as
\begin{align}
    \langle T^{00} \rangle \equiv \mathcal{U} (z) = \mathfrak{U} + f (z) . \label{T00Local}
\end{align}
We have relabeled the energy density for later use. So, the only part of the vacuum energy corresponding to an observable force is that coming from the first term, $\mathfrak{U}$, since the $z$-dependent term, $f (z)$, produces another divergent constant term and as such it does not contribute to the pressure. This can be easily confirmed by integrating it over $z$:
\begin{align}
    \int _{0} ^{L} f(z) dz = - \frac{1}{48 \pi ^{2}} \sqrt{\frac{1 - \lambda u _{z} ^{2}}{1 + \lambda u ^{2}}} \int _{2 m} ^{\infty} \sqrt{x ^{2} - 4 m ^{2}} ( x ^{2} + 2 m ^{2} ) \frac{dx}{x} .
\end{align}
Therefore the function $f(z)$ describes the local behavior of the energy density. In the massless case, the function $f (z)$ can be expressed in terms of the Hurwirtz zeta function, $\zeta (s,a) = \sum _{n = 0} ^{\infty} (n+a) ^{-s}$, as follows:
\begin{align}
    f (z) &= - \frac{1}{16 \pi ^{2} L ^{4}} \frac{(1 - \lambda u _{z} ^{2}) ^{2}}{\sqrt{1 + \lambda u ^{2}}} \left[ \zeta (4, z/L) + \zeta (4, 1 - z/L) \right] , \label{fz-massless}
\end{align}
which differs from the expression for the Lorentz-symmetric case only by the overall factor which depends on the Lorentz-violating parameters $\lambda$ and $u _{\mu}$. It is clear from the Eq. (\ref{fz-massless}) that the energy density for a massless scalar field diverges quartically as $z$ approaches to the plates. For a massive scalar field it is not clear the degree of divergence (due to the intricate structure of the integral in Eq. (\ref{fz})). In fig. \ref{LocalEffectsPlot} we show the singular part of the local energy density for $mL=1$ and $u _{0} ^{2} = u _{\perp} ^{2}$. The black-continuous line corresponds to the Lorentz symmetric case ($\lambda u _{z} = 0$), while the blue-dashed and red-dotted lines correspond to Lorentz-violating cases with $\lambda u _{z} = 0.5$ and $\lambda u _{z} = 0.95$, respectively. Our results indicate that there are no local effects when $\lambda u _{z} = 1$, independently of the values of $u _{0}$ and $u _{\perp}$. Indeed, this is clear from Eq. (\ref{fz-massless}) for the massless case.

Next we turn to the remaining components of the stress-energy tensor. From the rotational invariance around the $z$-axis, the components of the stress perpendicular to $n ^{\mu}$, $\langle T^{11} \rangle$ and $\langle T^{22} \rangle$, are equal. In addition, from the mathematical structure of the vacuum stress (\ref{VEVTmunu}) we find the relation $\langle T^{11} \rangle = - \langle T^{00} \rangle$. This is clearly seen by writing explicitly the $11$-component of the VS and performing an analysis similar to that of Sec. \ref{GlobalCasSec}. The $33$-component was extensively discussed in Sec. \ref{ST}. Let us relabel this component as $\mathcal{P} = \langle T^{zz} \rangle$. In the Lorentz-symmetric case, those are the only nonzero components of the vacuum stress \cite{BrownMaclay, Milton, Deutsch&Candelas}. In the problem at hand, as suggested by the stress-energy tensor in Eq.(\ref{Tmunu}), also the $a 3$-components will not be zero, with $a=0,1,2$. This is of course a direct consequence of the Lorentz-symmetry breaking. The vacuum expectation value of these components can be easily calculated using the above procedures. We close this section with an expression for the vacuum stress-energy tensor:
\begin{align}
\langle T^{\mu \nu} \rangle = (\eta ^{\mu \nu} + n ^{\mu} n ^{\nu}) \, \mathcal{U} (z) +  n ^{\mu} n ^{\nu} \, \mathcal{P} (z) - (\eta ^{\mu \alpha} + n ^{\mu} n ^{\alpha}) u ^{\alpha} n ^{\nu} \frac{2 \lambda u _{z}}{1 - \lambda u _{z} ^{2}} \left[ \mathfrak{U} - \mathcal{P} (z) \right] , \label{VSC}
\end{align}
where $n ^{\mu} = (0,0,0,1)$ is the normal to the plates. Clearly, in the limit $\lambda \to 0$, the last term vanishes and hence we recover the usual structure of the VS \cite{BrownMaclay}.

\section{Conclusions} \label{Conclusection}

In this paper we have considered a Lorentz-breaking extension of a real massive scalar quantum field theory. Such extension is described by the CPT-even aetherlike term $\lambda \left(u \cdot \partial \phi \right)^{2}$, where $\lambda$ is a dimensionless parameter and $u ^{\mu}$ is a background (constant) four-vector which control Lorentz symmetry breaking \cite{Gomes&Petrov}. Concretely, here we have analyzed the effects of Lorentz violation in the Casimir effect between two parallel conductive plates separated by a distance $L$. To this end, we have employed a field theoretical approach, based on Green's function techniques, which allow us to study the properties of the vacuum from the behavior of local field quantities. In the problem at hand, the stress-energy tensor $T ^{\mu \nu}$ represents the appropriate quantity, since $T ^{00}$ represents the local energy density, $T ^{0\mu}$ gives the flow of energy and momentum, and the stress components $T ^{ij}$ provide the mechanical properties of the vacuum \cite{Greiner}. 

A local formulation implies the introduction of the vacuum stress $\langle T ^{\mu \nu} \rangle$, i.e. the vacuum expectation value of the stress-energy tensor. Formally, the vacuum stress can be obtained by applying a certain second-order differential operator $\mathcal{O} ^{\mu \nu}$ upon the Green's function $G (x,x ^{\prime})$ and taking the limit $x ^{\prime} \to x$, i.e. $\langle T ^{\mu \nu} \rangle = \lim _{x ^{\prime} \to x} \mathcal{O} ^{\mu \nu} G (x,x ^{\prime})$. This is the so called point-splitting technique, which is possible given that the Green's function represents the vacuum expectation value of the time-ordered product of fields, i.e. $G (x,x ^{\prime}) = -i \langle 0 | \mathcal{T} \phi (x) \phi ( x ^{\prime}) |0\rangle$ \cite{BrownMaclay, Milton, Deutsch&Candelas}. In the problem at hand, the explicit form of the differential operator is read-off from Eq. (\ref{VEVTmunu}). In order to compute the vacuum stress between the plates, in Sec. \ref{Intro} we have derived in detail different Green's functions for the Lorentz-violating massive scalar field theory.

We first tackled the problem by integrating the energy density $\langle T ^{00} \rangle$ over $z$ in the region between the plates. This gives the Casimir energy, i.e. $\mathcal{E} _{C} (L) = \int _{0} ^{L} \langle T ^{00} \rangle \, dz$. We found that it can be expressed in a simple fashion in terms of the Lorentz-symmetric Casimir energy $\mathcal{E} _{0} (L)$ as
\begin{align}
    \mathcal{E} _{C} (L) = \sqrt{\frac{1-\lambda u _{z} ^{2}}{1 + \lambda u ^{2}}} \mathcal{E} _{0} (\tilde{L}) ,
\end{align}
where $\tilde{L} = L / \sqrt{1 - \lambda u _{z} ^{2}}$ is a rescaled length. This result is valid for both the massless and massive cases, for which $\mathcal{E} _{0} (L) = - \frac{ \pi ^{2}}{1440 L ^{3}}$ and $\mathcal{E} _{0} (L ) = - \frac{m ^{2}}{8 \pi ^{2} L} \sum _{n=1} ^{\infty} \frac{1}{n ^{2}} K _{2} (2mnL)$ \cite{Milton, Farina}, respectively. Clearly, our result correctly reduces to the standard case in the limit $\lambda \to 0$, as it should be. In fig. \ref{EnergyMassless} we show the Casimir energy as a function of $\sqrt{\lambda} u _{z} $, and we consider the cases in which $u ^{\mu}$ is: timelike, spacelike and lightlike. In all cases, the Casimir attraction is smaller than the Lorentz symmetric force. From a high-energy physics point of view, a slightly deviation is expected, since Lorentz-violating parameters are usually assumed to be small. In that case, the leading order contribution of the above result for $\lambda u _{z} ^{2} \ll 1$ can be derived through the coordinate redefinition method. Nevertheless, Lorentz-violating effective field theories also emerge in condensed matter systems, where the symmetry breaking parameters are not necessarily small and hence the coordinate redefinition method does not provide a good approximation. Instead, we have to employ nonpertubative methods, as the one used in this paper, to obtain the Casimir energy. As a result, we have obtained closed analytical expressions for arbitrary values of $\lambda$ and direction of $u ^{\mu}$ provided $\lambda u ^{2} _{z} < 1$ to avoid instabilities in the theory. Based on this, we think that condensed matter systems represents a possible arena to test our results. As a consistency check, we also derived an expression for the Casimir pressure by direct evaluation of the normal-normal component of the vacuum stress $\mathcal{F} _{C}$. The result, as expected, coincides with the one obtained by differentiating the Casimir energy, i.e. $\mathcal{F} _{C} = - \frac{\partial \mathcal{E} _{C}}{\partial L}$, and which indeed can also by written in terms of the Lorentz-symmetric force $\mathcal{F} _{0}$ as $\mathcal{F} _{C} (L) = \frac{1}{\sqrt{1 + \lambda u ^{2}}} \mathcal{F} _{0} (\tilde{L})$. 

We have also computed the general structure of the vacuum stress $\langle T ^{\mu \nu} \rangle (z)$, which is found to be nonsymmetric due to the presence of Lorentz violation. In particular, the time-time component $\mathcal{U} \equiv \langle T ^{00} \rangle (z)$ gave us the local behavior of the energy density between the plates. We found it can be written as the sum of two terms, $\mathcal{U} = \mathfrak{U} + f (z)$, where $\mathfrak{U} = \mathcal{E} _{C} /L$ gives the physical energy and the function $f(z)$ encodes the local effects (see Eq. (\ref{fz}) and fig. \ref{LocalEffectsPlot}). In the massless case, the function $f(z)$ takes a particularly simple form in terms of the Hurwitz zeta function, i.e. $f(z) \sim \frac{1}{L ^{4}} \left[ \zeta (4,z/L) + \zeta (4,1- z/L)  \right]$, which is quartically divergent as in the Lorentz-symmetric case. In the massive case, it is not clear the degree of divergence.

%{\color{red}We close this paper by commenting that in Ref. \cite{Petrov} the authors derived some particular expressions for the Casimir energy which correspond to different choices of the Lorentz-breaking four-vector $u ^{\mu}$. They computed the zero-point energy by summing over the modes, which is suitable for computing global quantities, but do not provide information about their local behavior. Therefore this work can be seen as complementary to that in Ref. \cite{Petrov}. The advantage of our results is that, since we have worked out the Casimir effect by nonperturbatively means, we have a more freedom in the parameter space ($\lambda$,$u^\mu$) to modify the standard results where Lorentz invariance is preserved.}

Our results can be generalized in different ways. For example, it is straightforward to calculate the Casimir effect between two parallel plates which satisfy Neumann or Robin boundary conditions. Also, it is possible to study the Casimir effect for different geometries, such as spherical and/or cylindrical conductive shells. Further, the temperature effects can be included straightforwardly. We leave these problems for future works.

\acknowledgements

A. M.-R. acknowledges support from DGAPA-UNAM Project No. IA101320. C. A. E. is supported by a UNAM- DGAPA postdoctoral fellowship and Project PAPIIT No. IN111518.

\appendix


\begin{thebibliography}{99}


\bibitem{Mocioiu} I. Mocioiu, M. Pospelov, and R. Roiban, Phys. Lett. B \textbf{489}, 390 (2000).

\bibitem{Carroll} S. M. Carroll, J. A. Harvey, V. A. Kosteleck\'y, C. D. Lane, and T. Okamoto, Phys. Rev. Lett. \textbf{87}, 141601 (2001).

\bibitem{Carlson} C. E. Carlson, C. D. Carone, and R. F. Lebed, Phys. Lett. B \textbf{518}, 201 (2001).

\bibitem{Anisimov} A. Anisimov, T. Banks, M. Dine, and M. Graesser, Phys. Rev. D \textbf{65}, 085032 (2002).

\bibitem{Burgess} C. P. Burgess, J. M. Cline, E. Filotas, J. Matias, and G. D. Moore, J. High Energy Phys. 03 (2002) 043.

\bibitem{Frey} A. R. Frey, J. High Energy Phys. 04 (2003) 012

\bibitem{Cline} J. M. Cline and L. Valc\'arcel, J. High Energy Phys. 03 (2004) 032.

\bibitem{Kostelecky} D. Colladay and V. A. Kosteleck\'y, Phys. Rev. D \textbf{55}, 6760 (1997).

\bibitem{Kostelecky2} D. Colladay and V. A. Kosteleck\'y, Phys. Rev. D \textbf{58}, 116002 (1998).

\bibitem{Casimir} H. B. G. Casimir, Proc. K. Ned. Akad. Wet. \textbf{51}, 793 (1948).

\bibitem{Sparnnaay} M. J. Sparnnaay, Physica (Amsterdam) {\bf24}, 751 (1958).

\bibitem{Poppenhaeger} K. Poppenhaeger, S. Hossenfelder, S. Hofmann, and M. Bleicher, Phys. Lett. B {\bf582}, 1 (2004).

\bibitem{Edery} A. Edery and V. N. Marachevsky, Journal of High Energy Physics {\bf12}, 035 (2008).

\bibitem{Cheng} H. Cheng, Mod. Phys. Lett. A {\bf 21}, 1957 (2006).

\bibitem{Quach} J. Q. Quach, Phys. Rev. Lett. {\bf114}, 081104 (2015).

\bibitem{Santos&Khanna-Grav} A. F. Santos and F. C. Khanna,Int J Theor Phys {\bf55}, 5356 (2016).

\bibitem{Jiawei} J. Hu and H. Yu, Phys. Lett. B {\bf767}, 16 (2017).

\bibitem{Cortijo} A. G. Grushin and A. Cortijo, Phys. Rev. Lett. \textbf{106}, 020403 (2011).

\bibitem{Cortijo2} A. G. Grushin, P. Rodriguez-Lopez, and A. Cortijo, Phys. Rev. B \textbf{84}, 045119 (2011).

\bibitem{MUC} A. Mart\'in- Ruiz, M. Cambiaso, and L. F. Urrutia, Europhys. Lett. \textbf{113}, 60005 (2016).

\bibitem{Kharlanov} O. G. Kharlanov and V. C. Zhukovsky, Phys. Rev. D \textbf{81}, 025015 (2010). 

\bibitem{CE&AM} A. Mart\'{i}n-Ruiz and C. Escobar, Phys. Rev. D \textbf{94}, 076010 (2016).

\bibitem{CE&AM2} A. Mart\'{i}n-Ruiz and C. Escobar, Phys. Rev. D \textbf{95}, 036011 (2016).

\bibitem{Frank&Turan} M. Frank and I. Turan, Phys. Rev. D \textbf{74}, 033016 (2006).

\bibitem{Santos&Khanna} A. F. Santos and F. C. Khanna, Phys. Lett. B {\bf762}, 283 (2016). 

\bibitem{Cruz&PetrovFermion} M. B. Cruz, E. R. Bezerra de Mello, and A. Yu. Petrov, Phys. Rev. D {\bf99}, 085012 (2019).

\bibitem{Santos&Khanna2} A. F. Santos and F. C. Khanna, Eur. Phys. Lett. {\bf125}, 41002 (2019).

\bibitem{Blasone} M. Blasone, G. Lambiase, L. Petruzziello, and A. Stabile, Eur. Phys. J. C {\bf78}, 976 (2018).

\bibitem{Gomes&Petrov} M. Gomes, J. R. Nascimento, A. Yu. Petrov, and A. J. da Silva, Phys. Rev. D \textbf{81}, 045018 (2010). 

\bibitem{Cruz&Petrov} M. B. Cruz, E. R. Bezerra de Mello, and A. Yu. Petrov, Phys. Rev. D \textbf{96}, 045019 (2017).

\bibitem{Altschul1} A. Ferrero and B. Altschul, Phys. Rev. D \textbf{84}, 065030 (2011).

\bibitem{Altschul2} B. Altschul, Phys. Lett. B \textbf{639}, 679 (2006).

\bibitem{Schwinger} J. Schwinger, L. DeRaad, K. Milton and W. Tsai, \textit{Classical Electrodynamics}, Advanced Book Program (Perseus Books, Boulder, 1998).

\bibitem{BrownMaclay} L. S. Brown and G. J. Maclay, Phys. Rev. \textbf{184},1272 (1969).

\bibitem{Milton} K. A. Milton, \textit{The Casimir effect: Physical Manifestation of Zero-Point Energy} (World Scientific, Singapore, 2001).


\bibitem{Schwinger2} J. Schwinger, Phys. Rev. {\bf82}, 664 (1951).

\bibitem{Schwinger3} J. Schwinger, Phys. Rev. {\bf94}, 1362 (1954).

\bibitem{Weisskopf} V. Weisskopf, Kgl. Danske Vid. Selskab {\bf14}, 166 (1936) .

\bibitem{Heisenberg} W. Heisenberg and H. Euler, Z. Phys. {\bf98}, 714 (1936).

\bibitem{Meara} O. T. O'Meara, \textit{Introduction to Quadratic Forms} ( Berlin, New York: Springer-Verlag, 2000).

\bibitem{Farina} C. Farina, Braz. J. Phys. \textbf{36}, 1137 (2006).

\bibitem{Deutsch&Candelas} D. Deutsch and P. Candelas, Phys. Rev. D {\bf20}, 3063 (1979).

\bibitem{Greiner} G. Plunien, B. M\"{u}ller, and W. Greiner, Phys. Rep. (Review Section of Physics Letters) {\bf132}, 87 (1986).




\end{thebibliography}
\end{document}